\documentclass[aps,twocolumn,prl]{revtex4}

\input epsf
\def\plotone#1{\centering \leavevmode
\epsfxsize= 1.0\columnwidth \epsfbox{#1}}

\def\apjl{Astrophys. J. Lett.}
\def\mnras{Mon.Not.Roy.As.Soc.}

\def\be{\begin{equation}}
\def\ee{\end{equation}}
\def\bea{\begin{eqnarray}}
\def\eea{\end{eqnarray}}

\def\Mpc{\,{\rm Mpc}}

\def\cmm2{{\,\rm cm^{-2}}}
\def\cm2{{\,{\rm cm}^2}}
\def\cmm3{{\,{\rm cm}^{-3}}}
\def\gcmm3{{\,{\rm g\,cm^{-3}}}}

\def\fun#1#2{\lower3.6pt\vbox{\baselineskip0pt\lineskip.9pt
  \ialign{$\mathsurround=0pt#1\hfil##\hfil$\crcr#2\crcr\sim\crcr}}}
\def\C{{\cal C}}

\def\vec{\bf}

\def\p3m{P$^3$M}

\def\la{\mathrel{\mathpalette\fun <}}
\def\ga{\mathrel{\mathpalette\fun >}}
\def\fun#1#2{\lower3.6pt\vbox{\baselineskip0pt\lineskip.9pt
  \ialign{$\mathsurround=0pt#1\hfil##\hfil$\crcr#2\crcr\sim\crcr}}}

\def\clphi{\C_l^{\phi \phi}}
\def\clee{\C_l^{EE}}
\def\clte{\C_l^{TE}}
\def\cltt{\C_l^{TT}}

\def\cldd{\C_l^{dd}}

\def\cleeu{{\tilde \C}_l^{EE}}
\def\clteu{{\tilde \C}_l^{TE}}
\def\clttu{{\tilde \C}_l^{TT}}

\def\aap{Astron. \& Astrophys.}

\newenvironment{tablehere}{\def\@captype{table}}{}
\newcommand{\tableskip}{\\[-6pt]}

\begin{document}
\bibliographystyle{prsty}
\title{Determining Neutrino Mass from the CMB Alone}
\author{Manoj\ Kaplinghat, Lloyd\ Knox and Yong-Seon\ Song }
\affiliation{Department of Physics, One Shields Avenue\\
University of California, Davis, California 95616, USA}
\date{\today}

\begin{abstract}
Distortions of Cosmic Microwave Background (CMB) temperature and 
polarization maps caused by gravitational lensing, observable
with high angular resolution and high sensitivity, can be used to 
measure the neutrino mass.  Assuming two massless species and one with
mass $m_\nu$ we forecast $\sigma(m_\nu) = 0.15$ eV from the Planck
satellite and $\sigma(m_\nu)=0.04$ eV from observations with twice the
angular resolution and $\sim 20$ times the sensitivity. 
A detection is likely at this higher sensitivity since 
the observation of atmospheric neutrino oscillations require
$\Delta m_\nu^2 \ga (0.04{\rm eV})^2$.
\end{abstract}
  \maketitle

{\parindent0pt\it Introduction.} 
Results from the WMAP \cite{bennet03a} show the standard cosmological
model passing a highly stringent test.  With this spectacular success
of the CMB as a clean and powerful cosmological probe, and of the
standard model as a phenomenological description of nature, it is
timely to ask  what can be done with yet higher resolution and higher
sensitivity such as offered by the Planck instruments and beyond.  In
this {\it Letter} we mostly focus on neutrino mass determination, with 
brief discussion of other applications. 

Eisenstein et al. \cite{eisenstein99} found that the Planck 
satellite can measure neutrino mass with an error of 0.26 eV.  
This sensitivity limit is related to the temperature at which
the plasma recombines and the photons last scatter off of the
free electrons, $T_{\rm dec}\simeq 0.3$ eV.  Neutrinos with
$m_\nu \la T_{\rm dec}$ do not leave any imprint on the
last-scattering surface that would distinguish them from $m_\nu=0$.   

Neutrinos with mass $m_\nu \la T_{\rm dec}$ would affect the amplitudes
of gravitational potential peaks and valleys at intermediate
redshifts. Massive neutrinos can collapse into potential wells when
they become non-relativistic, while massless ones freely stream out.
The observed galaxy power spectrum (which is
proportional to the potential power spectrum at sufficiently large
scales), combined with CMB observations can be
used to put constraints on $m_\nu$ \cite{hu98a}. At present such an
analysis yields an upper bound on $m_\nu$ of $\sim 0.3$ eV
\cite{spergel03}\footnote{No significant improvement is expected from
  combining Planck and the SDSS galaxy power 
spectrum \cite{hu98,eisenstein99}}.

The alteration of the gravitational potentials at late times changes
the gravitational lensing of CMB photons as they traverse these
potentials \cite{seljak96,bernardeau97}. Including the gravitational
lensing effect, we find that the Planck error forecast improves to
0.15 eV.  We also show that more ambitious CMB experiments can reduce
this error to $\sim 0.04$ eV.
These mass ranges are interesting because the atmospheric neutrino
oscillations require that at least one of the active neutrinos have 
$m_\nu > 0.04$ to 0.1 eV. More detailed considerations \cite{beacom02}
show that the sum of the active neutrino masses (which is what the CMB
is most sensitive to) should be at least 0.06 eV. 

Tomographic observations of 
the galaxy shear due to gravitational lensing can achieve
sensitivities to $m_\nu$ similar to what we 
find here \cite{hu99,abazajian02b}.  
Our work is distinguished by its sole reliance on CMB temperature
and polarization maps which have different potential sources of
systematic error.  Complementary techniques are valuable since both
of these will be very challenging measurements.


\newcommand{\bbnonu}{$\beta\beta 0 \nu$}

The most stringent laboratory upper bound on neutrino mass comes from
tritium beta decay end-point experiments \cite{tritium} which
limit the electron neutrino mass to $\la 2$ eV. 
Proposed exeriments plan to reduce this limit by one to two
orders of magnitude by searching for neutrinoless double beta decay
(\bbnonu) \cite{zdesenko03}.  A Dirac mass would elude this search,
but theoretical prejudice favors (and the see-saw mechanism requires)
Majorana masses. Like the CMB and galaxy shear observations, these
future \bbnonu \ experiments will be extremely challenging. 


{\parindent0pt\it Lensing of the CMB.} 
The intensity and linear polarization of the CMB are completely 
specified by the Stokes
parameters, $I$, $Q$ and $U$ which are related to the
unlensed Stokes parameters (denoted with a tilde) by 
$X({\vec n})=\tilde X({\vec n}+{\vec \delta n})$
where $X$ stands for $I$, $Q$ or $U$.
The deflection angle, ${\vec \delta n}$, is the
tangential gradient of the projected gravitational potential, 
\be
\phi({\vec n})= 2\int dr\Psi(r{\vec \hat n},r)(r-r_s)/(r r_s)\,, 
\ee
where $r$ is the coordinate distance along our past light cone, $s$
denotes the CMB last--scattering surface, ${\vec \hat n}$ is the unit
vector in  the ${\vec n}$ direction and $\Psi$ is the
three-dimensional gravitational potential.

The statistical properties of the $I$, $Q$ and $U$ maps are most simply
described in the transform space: $a_T({\vec l})$, $a_E({\vec l})$,
and $a_B({\vec l})$ where $a_T$ is the spherical harmonic transform of $I$
and $a_E$ and $a_B$ are the curl--free and gradient--free
decompositions, respectively, of $Q$ and $U$
\cite{kamionkowski97,seljak97}.   
In this transform space
the effect of lensing by mode $\phi({\vec L})$ (harmonic transform of
$\phi({\vec n})$) is to shift power from,
e.g., $\tilde a_T({\vec L - l})$ to $a_T(\vec l)$.  Lensing also mixes
$\tilde a_E$ into $a_B$ and any $\tilde a_B$ into $a_E$
\cite{zaldarriaga98b}, thus generating scalar B (curl) mode
correlations.

\begin{figure}[htbp]
  \begin{center}
    \plotone{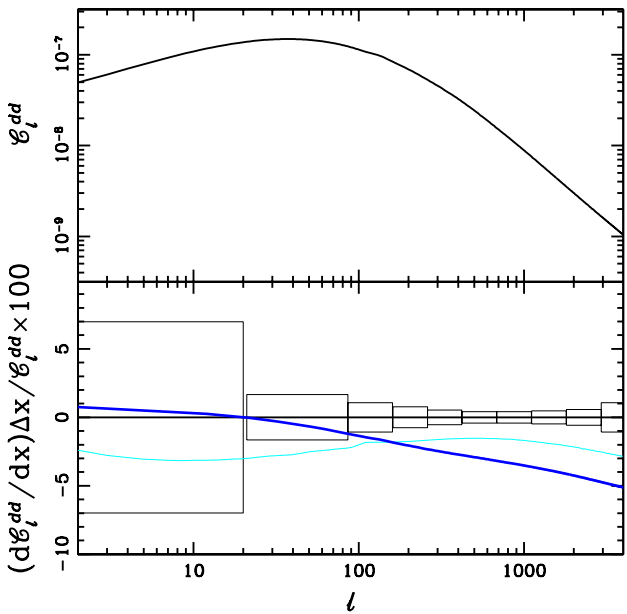}
    \caption{Top panel:  Deflection angle power spectrum $\cldd$
      for the fiducial model ($m_\nu = 0$). Bottom panel:  
$100 \times d \C_l^{dd} / d m_\nu \times (\Delta m_\nu/\C_l^{dd})$
(dark) and 
$100 \times d \C_l^{dd} / d w_x \times (\Delta w_x/\C_l^{dd})$ (light)
for  $\Delta m_\nu = 0.1$ eV and  $\Delta w_x = 0.2 $. 
\label{fig:cldd}
}
\end{center}
\end{figure}

Lensing smoothes out the features in the two-point functions, also
called angular power spectra, $\C^{\alpha \alpha'}_l$, where
$\langle a_\alpha({\vec l})a^*_{\alpha'}({\vec l'})\rangle =  
\C_l^{\alpha \alpha'} \delta({\vec l}-{\vec l'})/[2\pi l(l+1)]$ and
$\alpha$ stands for $T$, $E$, or $B$ \cite{seljak96}. As
explained later, in our analysis we use the unlensed power
spectra, ${\tilde \C}^{\alpha \alpha'}_l$. The information from
lensing is added through the two-point function of the lensing
potential, $\langle \phi({\vec L})\phi^*({\vec L'})\rangle = 
\C_L^{\phi \phi} \delta({\vec L}-{\vec L'})/[2\pi L(L+1)]$, 
which can be inferred from the temperature and
polarization map 4-point functions \cite{hu02b}. In Figure
\ref{fig:cldd} we plot the deflection angle power spectrum, 
$\cldd \equiv l(l+1) \clphi$.  


We calculate the 2-point functions using a publicly available code, 
CMBfast \cite{seljak96}, which was modified to include a scalar field
dark energy component, to calculate $\cldd$, and to include
the effect of massive neutrinos on the recombination history (through 
the expansion rate). We use the Peacock and Dodds prescription to
calculate the non-linear matter power spectrum \cite{peacock96}.  

{\parindent0pt\it Effect of neutrinos.} 
The lower panel in Figure \ref{fig:cldd} shows the differences in the
power spectra between our fiducial model and the exact same model but
with one of the three neutrino masses altered from zero to 0.1 eV.
The error boxes are those for CMBpol (described below; see Table
1). The $\cldd$ are noise-dominated at $l>600$ for CMBpol.  

The signature of a 0.1 eV neutrino in the angular power spectra,
in the absence of lensing, is at the 0.1\% level.
Such small masses are only detectable through their
effect on lensing, which comes through their influence on the
gravitational potential.  Replacing a massless
component with a massive one increases the energy density and
therefore the expansion rate, suppressing growth. The net suppression
of the power spectrum is scale dependent and the relevant length scale
is the Jeans length for neutrinos \cite{bond83,ma96,hu98} which
decreases with time as the neutrino thermal velocity decreases.
This suppression of growth is ameliorated at scales larger than the
Jeans length at matter--radiation equality, where the neutrinos can  
cluster. 
Neutrinos never cluster at scales smaller than the Jeans length
today. The net result is no effect on large scales and a suppression
of power on small scales, resulting in the shape of 
$\delta \C_l^{dd}/\C_l^{dd}$ in Figure \ref{fig:cldd}.  



{\parindent0pt\it Error forecasting method.} 

The power spectra we include in our analysis are 
$\clttu$, $\clteu$, $\cleeu$ (unlensed), and $\cldd$. 
We do not use the lensed power spectra to avoid the complication of
the correlation in their errors between different $\ell$ values and
with the error in $\cldd$. Using the lensed spectra and neglecting
these correlations can lead to overly optimistic 
forecasts \cite{hu02a}. If we include the lensed spectra instead of 
the unlensed ones, the expected errors on $w_x$ and $m_\nu$ for CMBpol
(see Table 1) shrink by about 40\% and 30\% respectively.

The distortions to the angular power spectra due to a $0.1$ eV
neutrino and changes of order 10\% in $w_x$ are very small.  We have
taken care to accurately forecast the constraints possible in this
mass range.  
First, we make a Taylor expansion of the power spectra to first order
in all the cosmological parameters. Then, given the the expected
experimental errors on the power spectra, the expected parameter error
covariance matrix is easily calculated.

The Taylor expansion works better and susceptibility to
numerical error is reduced with a careful choice of the parameters
used to span a given model space 
\cite{eisenstein99,efstathiou99,hu01a,kosowsky02}.  
We take our set to be ${\cal P} = \{\omega_m, \omega_b, \omega_\nu,
\theta_s, w_x,z_{\rm ri},k^3P_\Phi^i(k_f),n_s,n_s',y_{\rm He}\}$,
with the assumption a flat universe. 
The first three of these are the densities today (in units of
$1.88\times 10^{-29}{\rm g}/{\rm cm}^3$) of cold dark matter plus
baryons, baryons and massive 
neutrinos. Next two are the angular size subtended by the sound
horizon on the last--scattering 
surface and the ratio of dark energy pressure to density.   
The Thompson scattering optical depth for CMB photons, $\tau$, is
parameterized by the redshift of reionzation $z_{\rm ri}$.  
The primordial potential power spectrum is assumed to be
$k^3P_\Phi^i(k) = k_f^3P_\Phi^i(k_f)(k/k_f)^{n_s -1+n_S'ln(k/k_f)}$
with $k_f = 0.05\Mpc^{-1}$.   
The fraction of baryonic mass in Helium 
is $y_{\rm He}$.  We Taylor expand about
${\cal P}=\{0.146,0.021,0,0.6,-1,6.3,6.4\times 10^{-11},1,0,0.24\}$.

We follow \cite{zaldarriaga97} to calculate the errors expected in
$\clttu$, $\clteu$ and $\cleeu$ given Table 1.
For errors on $\cldd$ we follow \cite{hu02b}. 
The errors on the unlensed spectra in the regime where lensing is
important (deep in the damping tail) are certainly underestimated
because reconstruction of the unlensed map from the lensed map will
add to the errors. However, this is not worrisome since we limit all
the unlensed spectra to $l < 2000$, and a further restriction 
to $l < 1500$ (where lensing is least important) only increases the
error on $m_\nu$ by about 10\% for CMBpol.

{\parindent0pt\it Experiments.}  
We consider Planck \cite{planck}, a high-resolution version of 
CMBpol \footnote{CMBpol:
http://spacescience.nasa.gov/missions/concepts.htm.}, 
and a polarized bolometer array on the
South Pole Telescope \footnote{SPT:  http://astro.uchicago.edu/spt/} we
will call SPTpol.  Their specifications are given in Table 1.  
We assume that other frequency channels
of Planck and CMBpol (not shown in the table) will clean out non-CMB
sources of radiation perfectly.  Detailed studies have shown foreground
degradation of the results expected from Planck to be mild
\cite{knox99,tegmark00b,bouchet99}.  
At $l>3000$ emission from dusty galaxies will be a significant source
of contamination. The effect is expected to be more severe for
temperature maps. Hence we restrict temperature data to $l<2000$ and
polarization data to $l < 2500$. 

\begin{table}
\begin{center}
\begin{tabular}{ccccccc}
Experiment & $l_{\rm max}^{\rm T}$ &
$l_{\rm max}^{\rm E,B}$ & $\nu$ (GHz) & $\theta_b$ & $\Delta_T$ &
$\Delta_P$\\  
\tableskip\hline\tableskip
Planck    & 2000  & 2500 &  100 & 9.2' & 5.5 & $\infty$ \\
          &       &      &  143 & 7.1' & 6  & 11 \\
          &       &      &  217 & 5.0' & 13 & 27 \\
\tableskip\hline
SPTpol ($f_{\rm sky} = 0.1$)& 2000 & 2500 &  217 & 0.9' & 12 &
17 \\ 
\tableskip\hline
CMBpol      & 2000 &  2500 & 217 & 3.0' & 1  & 1.4 \\
\tableskip\hline
\end{tabular}
\end{center}
\caption{Experimental specifications. We use the unlensed spectra ($\clttu$,
  $\clteu$, $\cleeu$) only at $l < 2000$. For $\phi$
  reconstruction we use only data with $l<l_{\rm max}^{T,E,B}$. 
} 
\end{table}

\begin{tablehere}
\begin{table*}[hbt]\small
\caption{\label{table:bounds}}
\begin{center}
{\sc Error Forecasts}\\
\begin{tabular}{c|c|c|c|c|c|c|c|c|c|c}
\tableskip\hline\hline \tableskip Experiment & $m_\nu$ (eV) & $w_x$ &
$\ln P_\Phi^i$ & $n_S$ & $n_S'$ & $\theta_s$ (deg) & $\tau$ & $\ln
\omega_m$ & $\ln \omega_b$ & $y_{\rm He}$ \\
\tableskip\hline\tableskip
Planck   &  0.15&  0.31&  0.017&  0.0071&  0.0032&  0.002&  0.0088&
0.0066&  0.0075& 0.012\\ 
SPTpol   &  0.18&  0.49&  0.018&  0.01&  0.006&  0.0026&  0.0088&
0.0087&  0.01& 0.017\\ 
CMBpol  &  0.044&  0.18&  0.017&  0.0029&  0.0017&  0.00064&  0.0085&
0.0022&  0.0028 &  0.0048\\
\tableskip\hline
\end{tabular}\\[12pt]
\begin{minipage}{5.2in}
NOTES.---%
Standard deviations expected from Planck, SPTpol and CMBpol.
\end{minipage}
\end{center}
\end{table*}
\end{tablehere}

{\parindent0pt\it Results.} 
We emphasize the ability of
the experiments to simultaneously determine $P_\Phi^i$, $w_x$ and
$m_\nu$ \footnote{The degeneracy-breaking power of CMB lensing
was first pointed out for $\Omega_\Lambda/\Omega_k$ in
\cite{metcalf98}.}.   
These all affect the amplitude of $P_\Phi$ at late times, the 
latter two due to their effect on the rate of growth of density 
perturbations.  If we were only sensitive to the amplitude of 
$\cldd$ then there would be an exact degeneracy
between these three parameters.  However, the $l$-dependence of the
response of $\cldd$ to these parameter variations breaks this
would-be degeneracy, allowing for their simultaneous
determination. 

The effect of $m_\nu$ can easily be disentangled from that of $w_x$.  
We have already discussed 
the $l$-dependence of $\partial \ln \cldd / \partial m_\nu$ shown in
Fig. \ref{fig:cldd} as resulting from the scale- and time-dependence
of $\partial \ln P_\Phi/\partial m_\nu$.  
The $l$-dependence of $\partial \ln \cldd / \partial w_x$ has the
opposite sense.  Although the suppression of $P_\Phi$ for increasing
$w_x$ is nearly $k$--independent, the effect is larger at late times
---hence the radial projection gives a larger effect at low
$\ell$. The effects of $m_\nu$ and $w_x$ are sufficiently distinct to
allow for their simultaneous determination. We point out that the effect of
$w_x$ is more pronounced for larger values due to two reasons. One,
dark energy starts to dominate earlier (which implies larger uniform
suppression) and two, perturbations in dark energy on large scales are
enhanced for large $w_x$.  

The difference in the response of $\cldd$ to $m_\nu$ and $w_x$ 
allows for, e.g., Planck to detect the acceleration of the Universe
($w_x < -1/(3\Omega_x)$) at the 2$\sigma$ level.  Such a confirmation 
would be valuable given the deep theoretical implications of 
acceleration \cite{kaloper02}.  Hu \cite{hu02a} has previously noted
this result obtained with the assumption $m_\nu=0$. 

As is well known, the $P_\Phi^i$ can be determined independently
of the lensing signal, through use of a signal at large angular
scales.  One combines $\clee$ and $\clte$ at $l \la 20$ where they are 
proportional to $P_\Phi^i\tau^2$ and $P_\Phi^i\tau$ respectively 
\cite{zaldarriaga97,kaplinghat03a}
with the TT, EE and ET spectra at $20 \la l \la 2000$ where they are 
proportional to $P_\Phi^i e^{-2\tau}$.  

If we assume a single-step transition for the ionization history
Planck can achieve $\sigma(\tau)=0.005$ \cite{eisenstein99}.  However,
foreground contamination \cite{tegmark00b}, and modeling uncertainty
in the ionization history \cite{holder03} can increase this
uncertainty. For these reasons we conservatively ignore polarization
data at $l < 30$ and instead set a prior, by hand, of
$\sigma(\tau)=0.009$; including the $l < 30$ polarization data would
(perhaps artificially) achieve a smaller $\sigma(\tau)$.  In the end,
$\tau$ is determined (only slightly) better than this prior because
there is some constraint on $P_\Phi^i$ from the lensing signal.  Note
that since $P_\Phi^i e^{-2\tau}$ is so well-determined, we always
expect $\sigma(\ln P_\Phi) = 2 \sigma(\tau)$, as we find..

An extended period of reionization, as suggested by the combination
of WMAP and quasar observations \cite{kogut03}, may have
large spatial fluctuations in the ionization fraction.  Such
``patchy'' reionization would lead to a large diffuse kinetic SZ
contribution to $\cltt$ at high $l$
\cite{knox98,santos03}, possibly larger than the
lensing contribution. Fortunately the analogous effect in the
polarization is much smaller. 
For a conservative upper bound on how patchy reionization
could degrade $\sigma(m_\nu)$, we restrict the temperature data 
to $l < 1000$ and find $\sigma(m_\nu) = 0.045$ eV for CMBpol and
0.34 eV for Planck.


The primary motivation for
CMBpol is the detection of the B mode due to gravity waves produced
in inflation.  The amplitude of this signal would directly tell
us the energy density during inflation.  Following the calculation in
\cite{knox02, kesden02} we find a $3\sigma$ detection is possible for
CMBpol if the energy density during inflation is greater than
$\rho_{\rm min} = (2 \times 10^{15} {\rm GeV})^4$; 
$\rho_{\rm min}^{1/4}$ is an order of magnitude smaller than the GUT
scale. We note that $\rho_{\rm min} \propto 1/\tau$, approximately, for 
$0.05 < \tau < 0.2$ and we have assumed $\tau=0.1$.  This scaling with
$\tau$ suggests that the reionization feature in the B mode at the
largest angular scales is important and therefore a full-sky
experiment is necessary to achieve this sensitivity level. 


The scalar spectrum determined from high--resolution CMB observations
(the constraining power comes from primary CMB) 
can also be a useful probe of inflation, as studied recently by
\cite{gold03}.  If $n_S-1 = 0.07$, the central value in fits to WMAP
and other observations \cite{spergel03}, then inflationary models
generically predict $n_S' \sim (n_S-1)^2 = 0.005$ which will be
detectable at the $3 \sigma$ level by CMBpol.

Determining $\omega_b$ and $y_{He}$ to high precision will 
facilitate precision consistency tests with Big Bang Nucleosynthesis
(BBN) predictions. It will also be useful in
constraining non-standard BBN. For example, determining $\omega_b$ and
$y_{He}$ to high precision allows strong constraints to be put on
the number of relativistic species N (or equivalently the expansion
rate) during BBN. If $\sigma(y_{He})$ is small, then
$\sigma(N)=\sigma(y_{He})/0.013$, which for CMBpol works out to
$\sigma(N)=0.4$. Constraints on $N$ have important repercussions for
neutrino mixing in the early universe, and hence on neutrino mass
models \cite{abazajian02}. 

{\parindent0pt\it Conclusions.} 
Gravitational lensing of the CMB is a promising probe of the growth of
structure and the fundamental physics that affects it. High
sensitivity, high resolution  maps will allow us to measure the
lensing signature well enough to simultanteously constrain $m_\nu$,
$w_x$ and $P_\Phi$.  A future all-sky polarized CMB mission aimed at
detecting gravitational waves is likely to succeed in determining 
neutrino mass as well.   

\begin{acknowledgments}
We thank  J. Bock, S. Church, W. Hu, M. Kamionkowski, A. Lange,
S. Meyer and M. White for useful conversations.
\end{acknowledgments}

\end{document}